\documentclass[pra,twocolumn,showpacs,floatfix,amsfonts]{revtex4}
\usepackage{bm}
\usepackage{graphicx}
\usepackage{here}                    
\usepackage{latexsym}                
\usepackage{amsmath}
\usepackage{amssymb}  

\newcommand{\bra}[1]{\langle #1 | \,}
\newcommand{\ket}[1]{\, | #1 \rangle}

\newcommand{\be}{\begin{equation}}
\newcommand{\ee}{\end{equation}}
\newcommand{\bea}{\begin{eqnarray}}
\newcommand{\eea}{\end{eqnarray}}

\def\unity{\openone}

\begin{document}

\title{Robustness of the Bennett-Brassard 1984 quantum key 
distribution protocol against general coherent attacks}

\author{Georgios M. Nikolopoulos}
\affiliation{Institut f\"ur Angewandte Physik, Technische 
Universit\"at Darmstadt, 64289 Darmstadt, Germany}
\author{Gernot Alber}
\affiliation{Institut f\"ur Angewandte Physik, Technische 
Universit\"at Darmstadt, 64289 Darmstadt, Germany}

\date{\today}

\begin{abstract}
It is demonstrated that for the entanglement-based version of the
Bennett-Brassard (BB84) quantum key distribution protocol, 
Alice and Bob share provable entanglement if and only if the estimated 
qubit error rate is below $25\%$ or above $75\%$.
In view of the intimate relation between entanglement and security,
this result sheds also new light on the unconditional
security of the BB84 protocol in its original prepare-and-measure
form. In particular, it indicates that for small qubit error rates 
$25\%$ is 
the ultimate upper security bound for any prepare-and-measure 
BB84-type QKD 
protocol. On the contrary, for qubit error rates between $25\%$ and 
$75\%$
we demonstrate that the correlations shared between Alice and Bob can 
always be explained by separable states and thus, no secret key can be 
distilled in this regime.   
\end{abstract}

\pacs{03.67.Dd, 03.65.Ud}


\maketitle

\section{Introduction}
Quantum cryptography solves an outstanding problem of classical 
cryptography, namely the distribution of
a secret perfectly correlated random key between legitimate users. 
The first quantum key distribution (QKD) protocol, which was able to 
achieve this goal by
exploiting characteristic quantum phenomena, was proposed 
by Bennet and Brassard \cite{BB84}.
This BB84 protocol is the prototype of a prepare-and-measure QKD 
protocol, which allows unconditionally secure key distribution between 
two legitimate users, Alice and Bob. 
Its security against an eavesdropper, Eve, is guaranteed by 
the peculiar features of the quantum mechanical measurement process 
\cite{FGNP}. 
Later, it was realized that 
the security of the BB84 protocol is based on an effective 
entanglement shared between Alice and Bob. 
In particular, it was shown that, from the point of view of Eve, the 
originally proposed prepare-and-measure scheme 
is equivalent to an associated entanglement-based version of the BB84 
protocol \cite{BBM,LC,SP,GP}. 
The security of this latter protocol is guaranteed, for example, if 
the state of the entangled-qubit pairs shared between Alice and Bob 
is exponentially close to being a pure state. Then, according to 
Holevo's  theorem, Eve's mutual information with the distributed 
key is exponentially small \cite{GP,LC}. 

This equivalence between a prepare-and-measure QKD protocol and its 
associated entanglement-based QKD protocol, has been exploited in 
various proofs of unconditional security of the BB84 protocol 
\cite{LC,SP,GP}. 
Indeed, that quantum entanglement is 
necessary for secret QKD, was also emphasized in a more general 
context, recently \cite{CLL,AG}.
Nevertheless, though with respect to unconditional security issues the 
prepare-and-measure and the entanglement-based forms of the BB84  
protocol are equivalent, as far as realistic implementations are 
concerned both schemes are quite different.
Specifically, entanglement-based QKD protocols typically require 
quantum memories and quantum computers for performing the entanglement 
purification, necessary for enforcing secrecy even in the presence of 
an arbitrary powerful eavesdropper. On the contrary, 
the corresponding prepare-and-measure 
schemes do not require this advanced quantum 
technology, and are therefore realizable much more easily with present 
day technology.

An important quantity characterizing the security of a QKD protocol is 
the {\em threshold disturbance}, i.e. the maximal disturbance or
quantum bit error rate (QBER) which can be tolerated by Alice and Bob 
for being capable of producing a secret, perfectly correlated
random key. This maximal disturbance quantifies the robustness of a 
QKD scheme, and in general it depends not only
on the eavesdropping strategy, but also on the algorithm (distillation 
protocol) 
Alice and Bob are using for post-processing their raw data \cite{AGS}.
From the practical point of view, it is most interesting to know the 
threshold disturbance of a given QKD protocol, with respect
to the most general eavesdropping attacks which are consistent 
with the laws of quantum mechanics.
For such general coherent attacks, it has been demonstrated, for 
example, that the prepare-and-measure scheme of the
BB84 protocol is unconditionally secure up to QBERs of 
approximately
$20\%$, provided one uses appropriate error correction and 
privacy amplification protocols, which 
involve two-way classical communication \cite{C,GL}. 
Furthermore, it is also known that a simple intercept-resend attack 
can beat the BB84 protocol giving rise to 
${\rm QBERs} \geq 25\%$ \cite{IR}. 
However, so far it is unknown 
whether an arbitrary powerful eavesdropper, whose power is only 
limited by the general laws of quantum mechanics, is able to 
push the threshold disturbance below $25\%$ 
\cite{AGS,GL,RMP}.

In this paper we investigate the robustness of the BB84 protocol, 
with respect to
general coherent attacks by an adversary whose power is 
only limited by the fundamental laws of quantum mechanics.
As a main result, it is shown that for QBERs less than $25\%$ Alice 
and Bob always share pairs of entangled qubits, 
even if Eve has performed a general coherent attack involving quantum 
memories and quantum computers. In other words, as far as the 
robustness of the protocol is concerned, the coherent attacks do not 
seem to offer any advantage over collective or incoherent attacks. 
For the entanglement-based version of the BB84 protocol
our result implies that Alice and Bob can distribute a random secret 
key up to this QBER, provided they purify their corrupted pairs 
with the help of an entanglement purification protocol (EPP)
\cite{DEJ,BDSW}. This result also sheds new light onto the still open 
question of the maximal error rate that can be tolerated by a 
prepare-and-measure scheme of the BB84 protocol \cite{GL,RMP}. 
In particular, in view of the equivalence between the 
prepare-and-measure scheme and the entanglement-based version of the 
BB84 protocol, our result demonstrates that 
$25\%$ is indeed the threshold disturbance for the prepare-and-measure 
scheme as well. Thus, in principle Alice and Bob are able to 
exploit these quantum correlations for the distillation of a perfectly
correlated secret random key, by means of an appropriate classical 
distillation protocol which involves two-way classical communication 
\cite{C,GL}. 
Furthermore, as long as Alice and Bob focus on their sifted 
key and manipulate each qubit-pair independently, for QBERs between 
$25\%$ and $75\%$ a secret-key distillation is impossible. 
In particular, in this case we show that the correlations shared 
between Alice and Bob can always be described by separable states 
and thus the necessary precondition for secret-key 
distillation is violated \cite{CLL,AG}.   

This paper is organized as follows: For the sake of completeness 
in Sec. \ref{basics} we summarize briefly
basic facts of the prepare-and-measure scheme and the equivalent 
entanglement-based version
of the BB84 protocol, as far as they are significant for our 
subsequent discussion.
In Sec. \ref{n-coherent} we present the proof of our main result while 
Sec. \ref{conclusions} contains discussion and conclusions.

\section{BASIC FACTS ABOUT BB84 PROTOCOL}
\label{basics} 
In the prepare-and-measure BB84 protocol, 
Alice sends a sequence of qubits to Bob each of which is randomly 
prepared in one of the basis
states $ \{\ket{0},\ket{1}\}$ or $\{\ket{\bar 0},\ket{\bar 1}\}$ 
which are eigenstates of
two maximally conjugate physical variables, namely the two Pauli 
spin operators ${\cal Z}$ and ${\cal X}$.
The eigenstates of
${\cal Z}$, i.e. $\{\ket{0},\ket{1}\}$, and of 
${\cal X}$, i.e. $\{\ket{\bar 0},\ket{\bar 1}\}$, 
are related by the Hadamard transformation
\bea
{\cal H}=\frac{1}{\sqrt{2}}\left ( \begin{array}{cc} 
1 & 1 \\
1 & -1 \\
\end{array}\right ),
\eea
i.e.
$\ket{\bar i}=\sum_j {\cal H}_{ij}\ket{j} ~~(i,j = 0,1)$.
In the basis $ \{\ket{0},\ket{1}\}$, the Pauli spin operators are 
represented by the matrices
\bea
{\cal X}=\left ( \begin{array}{cc} 
0 & 1 \\
1 & 0 \\
\end{array}\right ),\,
{\cal Y}&=&\left ( \begin{array}{cc} 
0 & -i \\
i &  0\\
\end{array}\right ),\,
{\cal Z}=\left ( \begin{array}{cc} 
1 & 0 \\
0 & -1 \\
\end{array}\right ).
\eea 
Bob measures the  received  qubits randomly in one of the two bases.
Afterwards Alice and Bob apply a random permutation of their data, 
and publicly discuss the bases chosen, discarding all the bits where 
they have selected different bases. 
Subsequently, they randomly select half of the bits from the remaining 
random key (sifted key) and determine its error probability or QBER.
If, as a result of a noisy quantum channel or of an eavesdropper, the  
estimated QBER is too high, they abort the protocol.
Otherwise, they perform  error correction and privacy amplification  
with one- or two-way classical communication, in order to obtain a 
smaller number of secret and perfectly correlated random bits.

It has been shown that, from the point of view of 
an arbitrarily powerful eavesdropper, this originally proposed 
prepare-and-measure 
form of the BB84 protocol is equivalent to an entanglement-based 
QKD protocol \cite{BBM,LC,SP,GP} .
This latter form of the BB84 protocol offers advantages in particular 
with respect to questions concerning its unconditional security, and 
works as follows: 
Alice prepares each of, say  $2n$, entangled-qubit pairs in a 
particular Bell state \cite{bell}, say 
$\ket{\Phi^+}\equiv\frac{1}{\sqrt{2}}(\ket{0_A0_B}+\ket{1_A1_B})$ 
(where the subscripts $A,B$ refer to Alice and Bob, respectively). 
After having applied Hadamard transformation on the second half of 
each pair of a subset of randomly chosen qubit pairs, she sends all 
the second halves of the $2n$ pairs to Bob. 
Due to channel noise and the presence of a possible eavesdropper, 
however, at the end of this transmission stage all the $2n$ 
entangled-qubit pairs will be corrupted. 
In fact, they will be entangled among themselves as well as 
with Eve's probe. 

In order to ensure secret key distribution even in the presence of 
such a powerful eavesdropper, first of all Alice and Bob perform a 
{\em random permutation} of all the pairs, thus distributing any 
influence of the channel noise or the eavesdropper, equally among all 
the pairs \cite{GL,SP}. 
Afterwards, they perform a verification test, in which they 
randomly select half of the pairs \cite{half} as check pairs 
and measure each one of them {\em independently}, along a common 
basis, i.e. the quantization axis of the Bell state $\ket{\Phi^+}$.
However, these measurements are done only after Bob has reversed 
Alice's randomly applied Hadamard transformations.
For this latter purpose, after the transmission stage of the protocol,
Alice has to announce publicly the qubits on which she has applied 
Hadamard transformations.
Subsequently, the influence of channel noise or of an eavesdropper is 
quantified by the average
QBER of these $n$ check pairs, i.e. the probability with which Alice's 
and Bob's measurement results disagree. Note that, since Alice and Bob 
perform their measurements along the quantization axis of the Bell 
state $\ket{\Phi^+}$, in the absence of noise and eavesdropping their 
measurement results would perfectly agree. Furthermore, 
assuming that the check pairs constitute a fair sample \cite{half}, 
one may conclude that the same QBER applies also to the remaining, yet 
unmeasured, $n$ pairs.
After this verification test all the check pairs are dismissed and, 
if the QBER is too high, the protocol is aborted. 
Otherwise, Alice and Bob proceed to the purification of the 
remaining $n$ pairs

For this purpose, they apply an appropriate entanglement purification 
protocol (EPP) 
\cite{DEJ,BDSW} with classical one- or two-way communication,
so that they are able to distill a smaller number of almost pure  
entangled-qubit pairs. Finally, measuring these  almost perfectly 
entangled-qubit pairs in a common basis, i.e. along the quantization 
axis of the Bell state $\ket{\Phi^+}$, 
Alice and Bob obtain a perfectly correlated random key, 
about which an adversary has negligible information. 
For the sake of completeness we finally want to mention that this 
entanglement-based protocol can be reduced 
to an associated prepare-and-measure BB84 protocol with error 
correction and privacy amplification involving one- or two-way 
classical communication, only if the underlying EPP fulfills certain 
consistency conditions which are of no concern
for our further discussion \cite{SP,C,GL}. 

On the basis of the equivalence of these two forms of the BB84  
protocol, it has been demonstrated that by means of one-way classical 
distillation protocols a random key can be exchanged secretly 
between the two legitimate users up to a QBER of approximately 
$11\%$ \cite{SP}. Furthermore, post-processing that involves 
two-way classical communication can guarantee the extraction of 
a secret key even up to QBERs of the order of $20\%$ \cite{C,GL}.
On the other hand, it is also known that an intercept-resend attack 
can always defeat the BB84 protocol (irrespective of the 
distillation protocol employeed by Alice and Bob) for QBERs above 
$25\%$ \cite{IR}.
However, as far as QBERs between $20\%$ and $25\%$ are concerned, 
it is still an open question whether Alice and Bob are able to 
exchange a secret key, under the assumption of general coherent 
attacks. 

\section{Entanglement and threshold disturbance of the BB84 protocol}
\label{n-coherent} 
According to a recent observation, 
a {\em necessary precondition} for secret key distillation 
is that the correlations established between Alice and Bob 
during the state distribution cannot be explained by a separable 
state \cite{CLL,AG}. Throughout this work, we consider that Alice and 
Bob focus on the sifted key during the post processing 
(i.e., they discard immediately all the polarization data for which 
they have used different bases) and that they treat each pair 
independently. 
Thus, according to the aforementioned precondition, given a particular 
value of the estimated QBER, the task of Alice and Bob is to infer 
whether they share provable entanglement or not. 
Thereby, entanglement is considered to be provable if Alice's and 
Bob's correlations cannot be explained by a separable state within the 
framework of the protocol considered.  
In this section we demonstrate for the BB84 protocol that, as long 
as the detected QBER is less than $25\%$ or larger than $75\%$, Alice 
and Bob can be confident that they share entangled-qubit pairs 
with high probability. Furthermore, for QBERs between $25\%$ and 
$75\%$ the correlations between Alice and Bob can always be explained 
by a separable two-qubit state. 

Adopting the entanglement-based version of the
BB84 protocol we described in the previous section, the average
disturbance that Alice and Bob estimate during the verification test 
is given by \cite{GL,SP,qber}
\begin{widetext}
\bea
{\rm QBER} = \frac{1}{2}\sum_{b=0,1}~\sum_{l=0,1}
~\frac{1}{n}\sum_{j_i;~ i=1}^{n}
{\rm Tr}_{A,B}\{
[{\cal H}_A^b 
|l\rangle_{AA}\langle l|
{\cal H}_A^b
\otimes 
{\cal H}_B^b
|l\oplus1\rangle_{BB}\langle l\oplus1|
{\cal H}_B^b]_{j_i}~ 
 \rho_{AB}
\},
\label{QBER-pro}
\eea
\end{widetext}
where $\oplus$ denotes addition modulo 2.
Thereby  $\rho_{AB}$ denotes
the reduced density operator of Alice and Bob for all $2n$ pairs
while the index $j_i$ indicates that the
corresponding
physical observable
refers to the $j_i$-th randomly selected qubit pair.
The powers of the Hadamard transformation ${\cal H}^b$, with
$b\in\{0,1\}$, reflect the fact that
the errors in the sifted key originate from measurements in
both complementary bases which have been selected randomly by Alice
and Bob with equal probabilities.

Recall now that Alice and Bob have permuted all their
pairs randomly before the verification stage.
This random permutation has the effect of distributing the errors
introduced by Eve, over all the qubit pairs in a homogeneous way
\cite{GL,SP}.
As a result of this {\em homogenization procedure}
all the reduced density operators describing the state of each pair
shared between Alice and Bob become equal, i.e.
\begin{eqnarray}
\rho_{AB}^{(1)} &=& \rho_{AB}^{(2)} =\cdots = \rho_{AB}^{(2n)}.
\label{homogen}
\end{eqnarray}
The reduced density operator of Alice's and Bob' s $k$-th
pair is denoted by
$\rho_{AB}^{(k)} = {\rm Tr}_{AB}^{(\not k)}(\rho_{AB})$ and
${\rm Tr}_{AB}^{(\not k)}$ indicates the tracing (averaging) procedure
over all the qubit pairs except the $k$-th one.
Note that in general $\rho_{AB}$ is expected to have a complicated
structure as it includes all the effects arising from a general
coherent attack of a possible eavesdropper. As a result of the
homogenization procedure, however, the reduced states of the pairs
$\rho_{AB}^{(k)}$ are equal and thus the average estimated disturbance
(\ref{QBER-pro}) reads 
\begin{widetext}
\begin{eqnarray}
{\rm QBER} =
\frac{1}{2}\sum_{b=0,1} \sum_{l=0,1}
{\rm Tr}_{A,B}^{(j_1)}
\{
[{\cal H}_A^b
|l\rangle_{AA}\langle l|
{\cal H}_A^b
\otimes
{\cal H}_B^b
|l\oplus 1\rangle_{BB}\langle l\oplus 1|
{\cal H}_B^b]_{j_1}
\rho_{AB}^{(j_1)}
\},
\label{QBER}
\end{eqnarray} 
\end{widetext}
where ${\rm Tr}_{A,B}^{(j_1)}$ denotes the tracing procedure over the
$j_1$-th qubit pair of Alice and Bob.
According to Eq.  (\ref{QBER}), and in view of the homogenization
procedure, the QBER is determined by the
{\em average error probability}
of an arbitrary qubit pair, say the pair $j_1$.
Thus, the corresponding reduced density operator $\rho_{AB}^{(j_1)}$
contains all the necessary information about the noisy quantum channel
and
about a possible general coherent attack
by an eavesdropper, which is relevant for the evaluation of
the average QBER.
In particular, this implies that an arbitrary attack by an
eavesdropper which gives rise to
a particular state $\rho_{AB}^{(j_1)}$ is indistinguishable, from the
point of view of the estimated average disturbance, from
a corresponding individual attack which results in a decorrelated
$n$-pair
state of the form $\bigotimes_{i=1}^{n}\rho_{AB}^{(j_i)}$.

In order to address the problem for which 
values of the QBER the reduced density operator $\rho_{AB}^{(j_1)}$
is entangled, we start from the observation that the QBER of 
Eq. (\ref{QBER}) is invariant under the transformations
\begin{eqnarray}
(l,b) &\to& (l\oplus 1,b),\nonumber\\ 
(l,b) &\to& (l,b\oplus 1).
\label{Symm1}
\end{eqnarray}
This invariance implies that there are various reduced density 
operators of the $j_1$-th qubit pair, which all give rise
to the same observed value of the QBER. 
This can be seen from the
elementary relations
\begin{eqnarray}
{\cal X}{\cal H}^b
 |l\rangle \langle l|
{\cal H}^b
{\cal X} &=& {\cal H}^b |l\oplus 1\oplus  b\rangle \langle l\oplus 1\oplus b|{\cal H}^b,\nonumber\\
{\cal Z}{\cal H}^b
 |l\rangle \langle l|
{\cal H}^b
{\cal Z} &=& {\cal H}^b |l\oplus b\rangle \langle l\oplus b|{\cal H}^b,\nonumber\\
{\cal X}{\cal Z}{\cal H}^b
 |l\rangle \langle l|
{\cal H}^b
{\cal Z}{\cal X} &=& {\cal H}^b \ket{l\oplus 1}\bra{l\oplus 1}{\cal H}^b.
\end{eqnarray}
Together with the invariance of the QBER under the 
transformations (\ref{Symm1}), these relations imply that
the reduced operators $\rho_{AB}^{(j_1)}$ and
\begin{eqnarray}
\tilde{\rho}_{AB}^{(j_1)} &=&\frac{1}{8}\sum_{g\in{\cal G}_1, h\in {\cal G}_2} U(h)U(g)\rho_{AB}^{(j_1)}U(g)^{\dagger}U(h)^{\dagger} 
\label{rhotilde}
\end{eqnarray}
give rise to the same value of the QBER. 
Thereby, the unitary and hermitian operators
\begin{eqnarray}
U(g_1) &=& {\cal X}_A\otimes {\cal X}_B,\nonumber\\
U(g_2) &=& {\cal Z}_A\otimes {\cal Z}_B,\nonumber\\
U(g_3) &=& -{\cal Y}_A\otimes {\cal Y}_B \equiv {\cal X}_A{\cal Z}_A\otimes {\cal X}_B{\cal Z}_B,\nonumber\\
U(g_4) &=& \unity_A\otimes \unity_B,
\end{eqnarray}
and
\begin{eqnarray}
U(h_1) &=& {\cal H}_A\otimes {\cal H}_B,\nonumber\\
U(h_2) &=& \unity_A\otimes \unity_B,
\end{eqnarray}
have been introduced, which
form  unitary representations of two discrete Abelian groups
${\cal G}_1 =\{g_1,g_2,g_3,g_4\}$ and ${\cal G}_2 =\{h_1,h_2\}$.
Invariance of $\tilde{\rho}_{AB}^{(j_1)}$ under these groups 
is induced by the symmetry transformations of Eq.(\ref{Symm1}) which 
leave the QBER invariant.

As a consequence of Eq. (\ref{rhotilde}), the density 
operator $\rho_{AB}^{(j_1)}$ is entangled if 
$\tilde{\rho}_{AB}^{(j_1)}$ is entangled, as both states are related 
by local unitary operations and convex summation.  
Hence, to determine the values of the QBER for which Alice and Bob 
share provable entanglement, it suffices to determine the QBERs 
for which the most general two-qubit state $\tilde{\rho}_{AB}^{(j_1)}$ 
(which is invariant under the Abelian discrete groups ${\cal G}_1$ and 
${\cal G}_2$) is entangled.  

Note now that, the hermitian operators $U(g_1)$ and 
$U(g_2)$ of the group ${\cal G}_1$
constitute already a complete set of commuting operators 
in the Hilbert space of two qubits. 
Their eigenstates are given by the Bell 
states \cite{bell}. 
Thus, the most general two-qubit state which is invariant under the 
Abelian 
group ${\cal G}_1$ is given by a convex sum of Bell states, i.e. 
\begin{widetext}
\bea
{\tilde \rho}_{AB}^{(j_1)}=&\lambda_{00}\ket{\Phi^+}\bra{\Phi^+}+
\lambda_{10}\ket{\Phi^-}\bra{\Phi^-}
+\lambda_{01}\ket{\Psi^+}\bra{\Psi^+}+
\lambda_{11}\ket{\Psi^-}\bra{\Psi^-}. 
\label{rhoAB-Bell}
\eea
\end{widetext}
Thereby, the non-negative parameters 
$\lambda_{\alpha \beta}$ 
have to fulfill the normalization condition 
\bea 
\sum_{\alpha,\beta\in\{0,1\}}\lambda_{\alpha\beta} = 1.
\label{norm}
\eea
Furthermore, additional invariance of the quantum state 
(\ref{rhoAB-Bell}) under 
the discrete group ${\cal G}_2$ implies that 
\bea
\quad \lambda_{01}=\lambda_{10}.
\label{const1}
\eea 
As a consequence of this last constraint and the normalization 
condition, only two of the four parameters 
$\lambda_{\alpha\beta}$ are independent.
According to Eq.(\ref{QBER}), the average QBER is related to these 
parameters by
\bea 
{\rm QBER} = \lambda_{11} + \lambda_{01}.
\eea

Let us now determine the possible values of the QBER for which Alice 
and Bob share provable entanglement. 
To this end, we can take advantage of the Peres-Horodecki criterion, 
for example, stating that the state ${\tilde \rho}_{AB}^{(j_1)}$ is 
disentangled {\em iff} its partial transpose is non-negative 
\cite{P-H}.
In our case, this condition is equivalent to the inequalities
\begin{eqnarray}
\lambda_{01} + \lambda_{11}\geq \mid \lambda_{00}-\lambda_{10}\mid,
\label{1}\\ 
\lambda_{00} + \lambda_{10}\geq \mid \lambda_{01}-\lambda_{11}\mid,
\label{2}
\end{eqnarray}

\begin{figure}[t]
\centerline{\includegraphics[width=7.5cm]{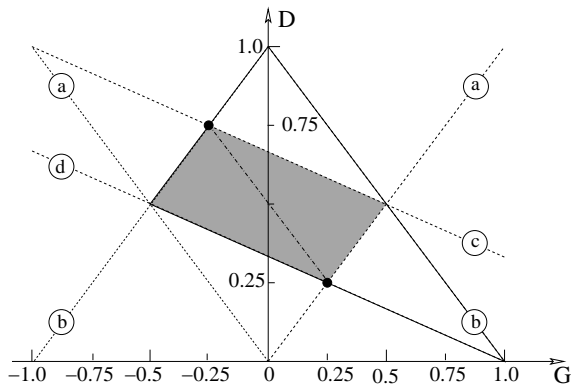}}
\caption{Regions of the independent parameters $D={\rm QBER}$ 
and $G = \lambda_{00}-\lambda_{10}$
for which the qubit pair state ${\tilde \rho_{AB}}^{(j_1)}$ is 
separable (shaded regions).
The various constraints that these parameters have to satisfy, are 
indicated by straight dashed lines, namely
(a) Eq. (\ref{1}); ~(b) $\lambda_{00}, \lambda_{10}\geq 0$;~
(c) Eqs. (\ref{norm}), (\ref{const1}) and (\ref{2}); ~
(d) $\lambda_{01}, \lambda_{11}\geq 0$ and Eqs. (\ref{norm}), 
(\ref{const1}). 
The protocol operates in the region which is defined by the solid 
lines. The threshold disturbances are indicated by black dots while  
the dot-dashed line connecting these dots corresponds to the family of 
separable states (\ref{sep}).
\label{D-G:fig}}
\end{figure}

As depicted in Fig. \ref{D-G:fig}, 
these inequalities, combined with Eqs. (\ref{norm}) and 
(\ref{const1}), imply 
that the reduced density operator ${\tilde \rho}_{AB}^{(j_1)}$ 
is entangled if the estimated QBER is below $1/4$ or above $3/4$. 
Given, however, that the states ${\tilde \rho}_{AB}^{(j_1)}$ 
and $\rho_{AB}^{(j_1)}$ are related via local operations 
and convex summation, the original single-pair 
state $\rho_{AB}^{(j_1)}$ must also be entangled in the same 
regime of parameters. 
Hence we may conclude that, whenever Alice and Bob detect an average 
QBER within this interval, they can be confident that 
they share entanglement with high probability, and their correlations 
cannot originate from a separable state.  The necessary precondition 
for 
secret-key distillation is therefore fulfilled. 

Let us now consider average QBERs with $1/4\leq {\rm QBER} \leq 3/4$. 
The fact that ${\tilde \rho_{AB}}^{(j_1)}$ is separable does not 
necessarily 
imply that $\rho_{AB}^{(j_1)}$ is also separable. But it indicates 
that 
in this regime of parameters, Alice's and Bob's correlations within 
the 
framework of the BB84 protocol can be explained by a separable state, 
namely 
by ${\tilde \rho_{AB}}^{(j_1)}$. According to \cite{CLL,AG}, this 
implies that 
Alice and Bob cannot extract a secret key. 
A particularly simple family of such separable states defined for 
$1/4\leq {\rm QBERs} \leq 3/4$, for example, is given by 
\begin{widetext}
\bea
\sigma_{AB}(D) = (4D-1)\frac{\unity_4}{4}+
\mid 1-2D\mid\sum_{k\in\{0,1\}} \left [ \frac{1}{2}
(\ket{k_A}\bra{k_A})\otimes
(\ket{k_B}\bra{k_B})+
\left ({\tilde \sigma}_{A}^{(k)}
\otimes {\tilde \sigma}_{B}^{(k)}\right)\right ].
\label{sep}
\eea
\end{widetext}
Thereby, $D\equiv{\rm QBER}$, $\unity_4$ denotes the unit 
operator in $\mathbb{C}_A^2\otimes\mathbb{C}_B^2$ and
${\tilde \sigma}_{C}^{(k)} =
\frac{1}{2}\left ( 
\ket{0_{C}}\bra{k_{C}}+\ket{1_{C}}\bra{(1\oplus k)_{C}} 
\right ).$

The family (\ref{sep}) is parameterized by the estimated average 
disturbance detected by Alice and Bob. 
Moreover, any separable state which belongs to this 
family is indistinguishable, from the point of view of the estimated 
QBER, from the real state shared between Alice and Bob. 
In other words, whenever the detected QBER is between 
$1/4$ and $3/4$, the correlations shared between Alice and Bob  
can be very well described in the framework of the family 
of separable states $\sigma_{AB}(D)$. The necessary precondition 
\cite{CLL,AG,AMG} 
for secret key distillation is not met for 
disturbances within this regime, so that the protocol must be aborted.  

\section{Discussion and Conclusions}
\label{conclusions}
We have discussed the threshold disturbance of the BB84 QKD protocol 
under the assumption of general coherent attacks which are only 
limited by the fundamental laws of quantum theory.
If Eve disentangles Alice and Bob by such a general attack 
(i.e., Alice's and Bob's correlations can be explained by a separable 
state) this gives rise to QBERs between $25\%$ and $75\%$. 
Hence, if Alice and Bob detect a QBER within this interval, they 
have to abort the protocol. As an example, we presented a particular 
class of separable states which is capable of describing the 
correlations shared between Alice and Bob for 
$25\%\leq {\rm QBER}\leq 75\%$. 
For QBERs below $25\%$ and above $75\%$ the two 
legitimate users can be confident that with high probability 
the quantum correlations they share cannot be described by separable 
states. 

The relevance of our results in the context of the entanglement-based 
version of the BB84 protocol is apparent : Alice and Bob 
are able to extract a secret  
key for ${\rm QBERs} < 25\%$ and ${\rm QBERs} > 75\%$ by application 
of an EPP which purifies towards the Bell states $\ket{\Phi^+}$ or  
$\ket{\Psi^-}$, respectively. 

However, for the prepare-and-measure forms of the BB84 protocol there 
are still open questions. 
So far, the highest tolerable error rates that have 
been reported in the context of general coherent eavesdropping attacks 
are close to $20\%$ \cite{C,GL}.  
Nevertheless, according to a recent observation \cite{CLL,AG}, a 
necessary precondition for secret-key distillation is that the 
correlations established between Alice and Bob during the state 
distribution cannot be explained by a separable state. 
In view of this strong link between quantum and secret correlations, 
our results suggest that in principle the distillation of a secret 
key is in principle possible for QBERs below $25\%$ and above $75\%$. 
In fact, if we focus on low disturbances, we can say that $25\%$  
is the upper bound on the maximal QBER that can be tolerated 
by any BB84-type prepare-and-measure QKD protocol. 
In other words, as far as the robustness of the BB84 protocol is 
concerned, coherent attacks do not seem to offer any advantages over 
collective or even incoherent attacks. Nevertheless, finding error 
correction and privacy amplification protocols required for a 
BB84-type protocol to be capable of meeting this upper security bound 
remains an open problem and will be addressed in subsequent work.

Finally, it should be stressed that the disturbance thresholds 
we have obtained depend on the post-processing of the BB84 QKD 
protocol. 
In particular, they rely on the complete omission of those qubits of 
the raw key for which Alice and Bob measured in different bases. 
Furthermore, they also rely on the fact that Alice and Bob manipulate
each qubit pair separately. At first sight it might appear to 
be puzzling why, under the same assumptions, it was shown that a 
secret key can be distilled in the framework of the BB84 protocol for 
QBERs up to about $30\%$, by means of advantage distillation 
protocols. These results, however, have been obtained under 
the additional assumption that Eve is restricted to so-called 
``optimal incoherent attacks'', i.e. attacks which maximize Eve's 
information gain. One can easily verify, for example, that the class 
of separable states (\ref{sep}) is not optimal (in the sense of 
\cite{AGS,GW}). In our work we allow for arbitrary eavesdropping 
attacks and thus we have demonstrated that the distillation of a 
secret key  for QBERs above $25\%$ is impossible \cite{CLL,AG,AMG}.  
So, although the incoherent attacks considered in \cite{AGS,GW} are 
optimal with respect to the information gain of an eavesdropper 
\cite{FGNP}, they are not able to disentangle Alice and Bob at the 
lowest possible QBER (that is at $25\%$). 
The cost of information loss that Eve has to accept by employing an 
attack that disentangles Alice and Bob at each particular disturbance 
between $25\%$ and $75\%$ remains an open question. 
Clearly, to this end one has to consider in detail the eavesdropping 
attack and this is beyond the purpose of this work. 
   
Nevertheless, in principle the distillation of a secret key for QBERs 
above $25\%$ is possible, provided that Alice and Bob do not 
focus on their sifted key only. In this context it was demonstrated 
with the help of entanglement witnesses recently, that the detection 
of quantum correlations between Alice and Bob is feasible even for 
QBERs above $25\%$, provided the data of the raw key are also 
taken into account \cite{CLL}. 

\section{Acknowledgments}
Stimulating discussions with
Norbert L\"utkenhaus are gratefully acknowledged. We thank an 
anonymous referee for insightful comments on the first version of 
this paper. This work is supported by the EU within the IP SECOQC.


\begin{thebibliography}{99}
\bibitem{BB84}
C. H. Bennett and G. Brassard, in {\em Proceedings of the 
IEEE International Conference on Computers, 
Systems and Signal Processing, Bangalore, India}
(IEEE, New York, 1984), pp. 175-179.

\bibitem{FGNP}
C. A. Fuchs, N. Gisin, R. B. Griffiths, C. S. Niu ,and A. Peres, 
Phys. Rev. A {\bf 56}, 1163 (1997).

\bibitem{BBM}
C. H. Bennett, G. Brassard, and N. D. Mermin, Phys. Rev. Lett. 
{\bf 68}, 557 (1992).

\bibitem{LC}
H.-K. Lo and H. F. Chau, Science {\bf 283}, 2050 (1999).

\bibitem{SP}
P. W. Shor and J. Preskill, Phys. Rev. Lett. {\bf 85}, 441 (2000).

\bibitem{GP}
D. Gottesman and J. Preskill, Phys. Rev. A {\bf 63}, 022309 (2001).

\bibitem{CLL}
M. Curty, M. Lewenstein, and N. L\"utkenhaus, Phys. Rev. Lett. 
{\bf 92}, 217903 (2003); see also quant-ph/0409047.

\bibitem{AG}
A. Acin and N. Gisin, quant-ph/0310054.

\bibitem{AGS}
A. Acin, N. Gisin, and V. Scarani, Quant. Info. Comp. {\bf 3}, 
563 (2003).

\bibitem{C}
H. F. Chau,  Phys. Rev. A {\bf 66}, 060302 (2002).

\bibitem{GL}
D. Gottesman and H.-K. Lo, IEEE Trans. Inf. Theory {\bf 49}, 
457 (2003).

\bibitem{IR}
C. H. Bennett, F. Bessette, G. Brassard, L. Salvail, and J. Smolin, 
J. Cryptology {\bf 5}, 3 (1992); A. Ekert and B. Huttner, J. Mod. Opt. 
{\bf 41}, 2455 (1994).

\bibitem{RMP}
N. Gisin, G. Ribordy, W. Tittel, and H. Zbinden, Rev. Mod. Phys. 
{\bf 74}, 145 (2002).

\bibitem{DEJ}
D. Deutsch, A. Ekert, R. Jozsa, C. Macchiavello, S. Popescu, and 
A. Sanpera, Phys. Rev. Lett. {\bf 77}, 2818 (1996).

\bibitem{BDSW}
C. H. Bennett, D. P. DiVincenzo, J. A. Smolin, and W. K. Wooters, 
Phys. Rev. A {\bf 54}, 3824 (1996); C. H. Bennett, Gilles Brassard, 
S. Popescu, B. Schumacher, J. A. Smolin and W. K. Wooters,  
Phys. Rev. Lett. {\bf 76}, 722 (1996).

\bibitem{bell}
The Bell states, 
$\ket{\Phi^\pm}\equiv\frac{1}{\sqrt{2}}(\ket{0_A0_B}\pm\ket{1_A1_B})$
and 
$\ket{\Psi^\pm}\equiv\frac{1}{\sqrt{2}}(\ket{0_A1_B}\pm\ket{1_A0_B})$, 
form an orthonormal basis in the two-qubit Hilbert space.

\bibitem{half}
It is worth noting that a logarithmic scaling of the size of the 
random sample with the length of Alice's and Bob's key, 
is sufficient for security issues. See Ref. \onlinecite{LCA} for a 
rigorous proof. 
Nevertheless, throughout this work we will consider that half 
of the pairs are sacrificed for the verification test.   

\bibitem{LCA}
H.-K. Lo, H. F. Chau and M. Ardehali, quant-ph/0011056.

\bibitem{qber}
Note that in the absence of noise and eavesdropping
each pair of qubits shared between Alice and Bob is
in the Bell state $\ket{\Phi^+}$. Thus, in this ideal system, 
Alice and Bob obtain perfectly correlated measurement results
whenever they perform their measurements along the same basis.

\bibitem{P-H}
A. Peres, Phys. Rev. Lett. {\bf 77}, 1413 (1996);
M. Horodecki, P. Horodecki, and R. Horodecki, Phys. Lett. A {\bf 223}, 
8 (1996). 

\bibitem{GW}
N.~Gisin and S.~Wolf, Phys. Rev. Lett. {\bf 83}, 4200 (1999). 
See also quant-ph/0005042.

\bibitem{AMG} 
A.~Acin, L.~Masanes, and N.~Gisin, Phys. Rev. Lett. {\bf 91}, 
167901 (2003).

\end{thebibliography}
\end{document}